\documentclass[12pt,prd,tightenlines,nofootinbib]{revtex4}
\usepackage{bm}
\usepackage{graphics}
\usepackage{rotating}
\usepackage{epsfig}
\usepackage{slashed,amsmath,float}
\usepackage{mathrsfs}
\usepackage{amsfonts}
\usepackage{graphicx}
\usepackage{color}
\usepackage{ulem,bbm}
\usepackage{subfigure}
\usepackage{diagbox}
\usepackage{lineno}
\usepackage{makecell}

\begin{document}
\title{
FCNC transitions of $\Lambda_b$ to neutron in Bethe-Salpeter equation approach}
\author{Liang-Liang Liu $^{a}$}
\email{corresponding  author. liu06_04@sxnu.edu.cn}
\author{Chao Wang $^{b}$}
\author{ Xian-Wei Kang $^{c}$}
\author{ Xin-Heng Guo $^{c}$}
\email{ corresponding  author. xhguo@bnu.edu.cn}

\affiliation{\footnotesize (a)~College of Physics and information engineering, Shanxi Normal University, Linfen 041004, People's Republic of China}
\affiliation {\footnotesize (b)~Center for Ecological and Environmental Sciences, Key Laboratory for Space Bioscience and Biotechnology, Northwestern Polytechnical University, Xi$^\prime$an, 710072, People's Republic of China}
\affiliation{\footnotesize (c)~College of Nuclear Science and Technology, Beijing Normal University, Beijing 100875, People's Republic of China}
\begin{abstract}
In a covariant quark-diquark model, we investigate the rare decay of $\Lambda_b \rightarrow n l^+ l^-$ and $\Lambda_b \rightarrow n \gamma$ in the Bethe-Salpeter equation approach.
In this model the baryons are treated as bound states of a constituent quark and a diquark interacting via a gluon exchange and the linear confinement.
We find that the ratio of form factors $R$ is varies from $-0.90$ to $-0.25$ and the branching ratios $Br(\Lambda_b \rightarrow n l^+ l^-)\times 10^{8}$ are $6.79(l=e),~4.08 ~ (l=\mu),~2.9 ~(l=\tau) $ and the branching ratio $Br(\Lambda_b \rightarrow \gamma)\times 10^{7} )=3.69$ in central values of parameters.
\end{abstract}

\maketitle

\section{Introduction}

The flavor changing neutral current (FCNC) decays of $b$-quark such as $b\rightarrow s \gamma ( l^+ l^-)$  can provide constrains on new physics, give essential information about the quark structure of heavy baryons and give more model-independent information such as CKM matrix elements.
Significant experimental progresses about rare decays of the $\Lambda_b$ baryon have been achieved at LHCb \cite{ PLB725-25, PRL123-031801, JHEP09-146, JHEP06-115}.
The rare decay $\Lambda_b \rightarrow \Lambda \mu^+ \mu^- $ first observed by CDF collaboration in 2011 \cite{PRL107-201802}.
The radiative decay $\Lambda_b \rightarrow \Lambda \gamma$ was observed at LHCb in 2019 \cite{PRL123-031801}.
There have been also many theoretical works on the rare decays $\Lambda_b$ induced by $b \rightarrow s$ transiton \cite{JPG24-979, PTP102-645, EPJC59-847, PRD87-074031, PRD96-053006, PTEP073B04, PRD59-114022,PRD53-4946, PRD63-114024, PLB-EPJC, EPJC-NPB, EPJC05-001, EPJC52-375, PRD81-056006, EPJC45-151, JHEP01-087, CTP-NPB, PRD93-074501, EPJC78-230,PLB-PLB}.
Ref. \cite{JPGNPP37-115007} gave the branching ratios $Br(\Lambda_b \rightarrow n l^+ l^-)\times 10^{8}= 3.19 \pm 0.46~(l=e),~3.76\pm 0.42~(l=\mu),~1.65\pm0.19~(l=\tau) $ in the context of light cone QCD sum rules (LCSR).
The form form factors (FFs) of $\Lambda_b (\Lambda_b^*) \rightarrow N l^+ l^-$ were given in Ref. \cite{PRD98-035033} in LCSR and taking into account the contribution of $\Lambda^*_b$ the ranching ratios  $Br(\Lambda_b \rightarrow Nl^+ l^-)\times 10^{8}= 8\pm2~(l=e),~7\pm2~(l=\mu),~ 2\pm0.4~(l=\tau) $ were obtained.
Ref. \cite{MPLA32-1750125} gave the branching ratios $Br(\Lambda_b \rightarrow n \mu^+ \mu^-)\times 10^{8}= 3.75\pm0.38 $ and $Br(\Lambda_b \rightarrow n \gamma)\times 10^{7}= 3.7 $  in the relativistic quark-diquark picture in the QCD-motivated interquark potential model.
Ref. \cite{EPJC79-383} studied the $\Lambda_b \rightarrow N^* l^+ l^-$ decay in LCSR and gave the branching ratios $Br(\Lambda_b \rightarrow N^* l^+ l^-)\times 10^{8}= 4.62\pm1.85~(l=e),~4.25\pm1.5~(l=\mu),$ and $0.25\pm0.09(l=\tau) $.
Ref. \cite{EPJC31-511} gave analysed CP-violation in polarized $b \rightarrow d l^+ l^-$.
With the experiment development, the transition $\Lambda_b \rightarrow n $ will be detected in the near future, so it is necessary to study $\Lambda_b \rightarrow n $ theoretically.

In this work, we will calculate the FFs of $\Lambda_b \rightarrow n $ in the Bethe-Salpeter equation approach in a covariant quark-diquark model.
This model has been used to study nucleon electromagnetic form factors and N-$\Delta$ transition form factors \cite{keiner}.
In the previous works, heavy baryon properties have been studied extensively in this model\cite{CPC42-103106, PRD95-054001, PRD54-4629, PRD87-076013, PRD91-016006, PLB954-97, PRD86-056006, PRD76-056004}.
The possible existence of diquark within baryons has been studied for a long time \cite{RMP65-1199, ZPC10-231, diquark-iii}.
The negative neutron mean square charge radius can be explained by diquark model, which cannot be explained in pure SU(6) quark model \cite{ZPC10-231}.

In our model, $\Lambda_b$ can be regarded as a bound state of two particles: one is a heavy quark $b$ and the other is a scalar diquark $(ud)$.
Using the $SU(6)$ wave function of baryons, we can get the neutron wave function in the quark-diquark model \cite{PR167-1523, PRD60-074017}.

\begin{eqnarray*}
  n^\uparrow &=& \frac{1}{3\sqrt{3}} [3 d^\uparrow (du)_{00}+d^\uparrow(du)_{10}-\sqrt{2}d^\downarrow(du)_{11}-\sqrt{2} u^\uparrow(dd)_{10}+2 u^\downarrow( dd)_{11}   ]
\end{eqnarray*}
where the arrow $\uparrow~(\downarrow)$ indicates the spin direction.
Therefore, in our model only the $d^\uparrow(ud)_{00}/ \sqrt{3}$ component of the neutron contributes to $\Lambda_b \rightarrow n$ since $\Lambda_b$ has the structure $b(ud)_{00}$.

This paper is organized as follows. In Section II, we will establish the BS equation for $q(ud)_{00}$ system ($q=b,~d$).
In Section III we will derive the FFs for $ \Lambda_b \rightarrow n$ in the BS equation approach.
In Section IV the numerical results for the decay FFs of $\Lambda_b \rightarrow n l^+ l^- $ will be given.
Finally, the summary and discussion will be given in Section V.

\section{BS EQUATION FOR $Q(ud)_{0 0}$ SYSTEM}\label{sec2}

Following our previous work, the BS equation of the $q(ud)_{0 0}$ system in momentum space satisfies the following homogeneous integral equation \cite{CPC42-103106, PRD95-054001, PRD54-4629, PRD87-076013, PRD91-016006, PLB954-97, PRD86-056006}:
\begin{eqnarray}\label{chi-p}
\chi_P(p) = i S_F(p_1)\int \frac{d^4 p}{(2 \pi)^4}  [ I\otimes I V_1(p,q)+ \gamma_\mu \otimes (p_2+q_2)^\mu  V_2(p,q)   ]\chi_P(q)S_D(p_2),
\end{eqnarray}
where $S_F(p_1)$ and $S_D(p_2)$ are propagators of the $q$ quark and the $(ud)$ scalar diquark, respectively, $p_1=\lambda_1 P+p$ and $p_2=\lambda_2 P-p$ correspond to the momenta of the quark and the diquark, respectively.
$P$ is the momentum of the baryon.
$V_1$ and  $V_2$ are the scalar confinement and one-gluon-exchange terms in the kernel, respectively.
Generally, the $q(ud)_{00}$ system needs two scalar functions to describe its BS wave function \cite{CPC42-103106, PRD95-054001, PRD91-016006}

\begin{figure}[!ht]
\begin{center}
\includegraphics[width=7.5cm] {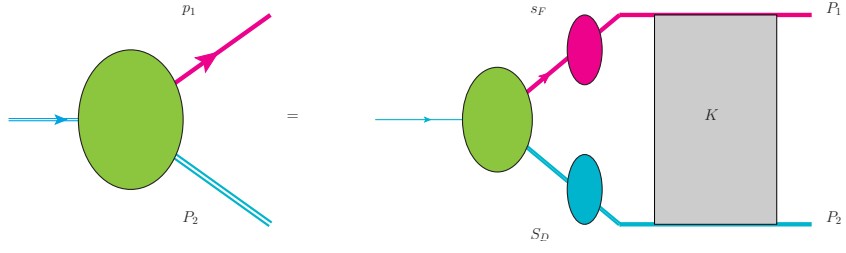}
\caption{The BS equation for the $q(ud)_{00}~(q=b,d)$ system in momentum space (K is the interaction kernel)}\label{BSE}
\end{center}
\end{figure}

\begin{eqnarray}
  \chi_P(p) &=& (f_1(p_t^2)+\slashed{p}_t f_2(p_t^2))u(P),
\end{eqnarray}
where $f_i~(i=1,2)$ are the Lorentz-scalar functions of $p_t^2$, $u(P)$ is the spinor of the baryons, $p_t$ is the transverse projection of the relative momenta along the momentum $P$, $p_t^\mu = p^\mu-(v\cdot p) v^\mu$, and $p_l= \lambda_2 M- v \cdot p  $ (where we defined $v^\mu=P^\mu/M$).
We use $M,~m,\text{and}~m_D$ to represent the masses of the baryons, the $q$-quark and the $(ud)$ diquark, respectively.

According to the potential model, $V_1$ and $V_2$ have the following forms in the covariant instantaneous approximation ($p_l=q_l$) \cite{PRD54-4629, PRD87-076013, PRD86-056006, PRD76-056004}:
\begin{eqnarray}\label{V1}
  \tilde{V}_1(p_t-q_t) = \frac{8 \pi \kappa}{[(p_t-q_t)^2+\mu^2]^2} - (2\pi)^2\delta^3(p_t-q_t)\int \frac{d^3k}{(2\pi)^3} \frac{8 \pi \kappa}{(k^2+\mu^2)^2},
\end{eqnarray}
where $q_t$ is the transverse projection of the relative momenta along the momentum $P$ and defined as $ q_t^\mu = q^\mu -(v \cdot q) v^\mu$, $q_l=\lambda_2 M -v \cdot q$. The second term of $\tilde{V}_1$ is introduced to avoid infrared divergence at the point $ p_t=q_t$, and $\mu$ is a small parameter to avoid the divergence in numerical calculations.
\begin{eqnarray}\label{V2}
  \tilde{V_2} (p_t-q_t)&=&- \frac{16 \pi }{3}\frac{\alpha^2_{seff} Q^2_0}{[(p_t-q_t)^2+\mu^2][(p_t-q_t)^2+Q_0^2]},
\end{eqnarray}
It was found that $Q_0^2=3.2$ GeV$^2$ can lead to consistent results with the experimental data by analyzing the electromagnetic FFs of the proton \cite{JPG24-979}.
The parameters $\kappa$ and $\alpha_{seff}$ are related to the scalar confinement and the one-gluon-exchange diagram, respectively.

The quark and diquark propagators can be written as the follows:
\begin{eqnarray}\label{SF}
  S_F(p_1) = i \slashed{v} \bigg[ \frac{\Lambda_q^+ }{ M -p_l -\omega_q  +i \epsilon} +\frac{\Lambda_q ^-}{ M -p_l +\omega  -i \epsilon}\bigg],
\end{eqnarray}
\begin{eqnarray}\label{SD}
  S_D(p_2)  = \frac{i}{2 \omega_D} \bigg[\frac{1}{ p_l-\omega_D+i \epsilon} -\frac{1}{ p_l+ \omega_D-i\epsilon}\bigg],
\end{eqnarray}
where $\omega_q = \sqrt{m^2-p_t^2}~\text{and}~\omega_D = \sqrt{m_D^2-p_t^2} $.
$\Lambda^\pm$ are the projection operators which have the following relations:
\begin{eqnarray}
2 \omega_q \Lambda^\pm_q&=& \omega_q \pm  \slashed{v}(\slashed{p}_t+m) , \\
\Lambda_q^\pm \Lambda_q^\pm &=& \Lambda^\pm_q,\\
\Lambda^\pm_q \Lambda^\mp_q &=& 0.
\end{eqnarray}

 Following our previous work, in order more precisely calculate the FFs of $\Lambda_b \rightarrow n$, we can take $E_0=-0.14$ GeV (where $E_0=M-m-m_D$ is the binding energy) and $\kappa$ to be about $0.05\pm0.005$ GeV$^3$ for $\Lambda_b\rightarrow n $ \cite{arXiv:1911.08023}.
Defining $\tilde{f}_{1(2)}=\int \frac{d p_l}{2 \pi}f_{1(2)}$, and using the covariant instantaneous approximation, $p_l=q_l$,  the scalar BS wave functions satisfy the following coupled integral equation:

\begin{eqnarray}\label{BS:f1}
&& \tilde{f}_1(p_t) =\int \frac{d^3q_t}{(2\pi)^3} M_{11}(p_t,q_t) \tilde{f}_1(q_t)+  M_{12}(p_t,q_t) \tilde{f}_2(q_t) ,
\end{eqnarray}

\begin{eqnarray}\label{BS:f2}
 &&\tilde{f}_2(p_t) = \int \frac{d^3q_t}{(2\pi)^3}  M_{21}(p_t,q_t) \tilde{f}_1(q_t) +  M_{22}(p_t,q_t) \tilde{f}_2(q_t),
\end{eqnarray}
where
\begin{eqnarray}
M_{11}(p_t,q_t)=\frac{(\omega_q  +m ) (\tilde{V}_1+ 2 \omega_D \tilde{V}_2)-   p _t \cdot ( p _t+ q _t) \tilde{V}_2}{4 \omega_D \omega_q(-M + \omega_D+ \omega_q)} - \nonumber\\ \frac{(\omega_q -m )(\tilde{V}_1- 2\omega_D \tilde{V}_2)+   p _t\cdot( p _t+ q _t)  \tilde{V}_2}{4 \omega_D \omega_c(M + \omega_D+ \omega_q)},
\end{eqnarray}

\begin{eqnarray}
M_{12}(p_t,q_t)=\frac{-  (\omega_q+m ) ( q _t + p _t)\cdot q_t\tilde{V}_2 +  p _t\cdot q_t(\tilde{V}_1- 2 \omega_D \tilde{V}_2)}{4 \omega_D \omega_c(-M + \omega_D+ \omega_c)}- \nonumber\\ \frac{(m - \omega_q )  ( q _t + p _t)\cdot q _t \tilde{V}_2 -   p _t\cdot q _t  (\tilde{V}_1+ 2\omega_D \tilde{V}_2)}{4 \omega_D \omega_q(M + \omega_D+ \omega_q)},
\end{eqnarray}

\begin{eqnarray}
M_{21}(p_t,q_t)= \frac{(\tilde{V}_1+ 2 \omega_D \tilde{V}_2)-( -\omega_q+m ) \frac{( p _t+ q _t) \cdot p _t }{ p^2_t }\tilde{V}_2}{4 \omega_D \omega_q(-M + \omega_D+ \omega_q)}    - \nonumber\\
\frac{- (\tilde{V}_1- 2\omega_D \tilde{V}_2)+(\omega_q  + m )\frac{( p _t+ q _t)\cdot p _t }{ p^2_t } \tilde{V}_2)}{4 \omega_D \omega_q(M + \omega_D+ \omega_q)},
\end{eqnarray}

\begin{eqnarray}
M_{22}(p_t,q_t)=  \frac{(m  -\omega_q)( \tilde{V}_1+ 2  \omega_D \tilde{V}_2）  ) \frac{ p_t \cdot q_t}{ p^2_t } - (  q^2_t+  p_t \cdot q_t) \tilde{V}_2}{4 \omega_D \omega_q(-M + \omega_D+ \omega_q)} - \nonumber\\
\frac{ (m +\omega_q) (-\tilde{V}_1- 2 \omega_D \tilde{V}_2）) \frac{p_t \cdot q_t}{p^2_t} + (  q^2_t+  p_t \cdot q_t)\tilde{V}_2)}{4 \omega_D \omega_q(M + \omega_D+ \omega_q)}.
\end{eqnarray}

When $ \frac{1}{m}\rightarrow 0$ \cite{PRD54-4629}, the quark propagator can be written as following,
\begin{eqnarray}\label{SF-HQ}
  S_F(p_1) = i \frac{ 1+ \slashed{v}  }{  2 (E_0+m_D -p_l+ i \epsilon) },
\end{eqnarray}
considering the Dirac equation for $\Lambda_b$ we have
\begin{eqnarray}\label{BS:hq}
  \phi(p) &=& -\frac{i}{(E_0+m_D-p_l+i \epsilon)( p_l ^2-\omega^2_D)}\int \frac{d^4 q }{(2\pi)^4}(\tilde{V}_1+2  p_l \tilde{V}_2)\phi(q),
\end{eqnarray}
where the BS wave function of $\Lambda_b$ was given in the previous work \cite{PRD54-4629} and has the form $\chi_P (v)=\phi (p)u_{\Lambda_b}(v,s)$ with $\phi(p)$ being the scalar BS wave function.

Generally, the BS wave function can be normalized under the condition of the covariant instantaneous approximation \cite{PRD76-056004}:
\begin{eqnarray}\label{BSNOR}
  i \delta^{i_1 i_2}_{j_1 j_2} \int \frac{d^4 q d^4 p}{(2\pi)^8}\bar{\chi}_P(p,s)\left[\frac{\partial}{\partial P_0}I_p(p,q)^{i_1 i_2 j_2 j_1}\right]\chi_P(q,s^\prime) =\delta_{s s^\prime},
\end{eqnarray}
where $i_{1(2)}$ and $j_{1(2)}$ represent the color indices of the quark and the diquark, respectively, $s^{(\prime)}$ is the spin index of the baryon, $I_p(p,q)^{i_1 i_2 j_2 j_1}$ is the inverse of the four-point propagator written as follows
\begin{eqnarray}\label{IPNOR}
  I_p(p,q)^{i_1 i_2 j_2 j_1} =\delta^{i_1 j_1}\delta^{i_2 j_2} (2 \pi)^4 \delta^4(p-q)S^{ -1 }_F(p_1)S^{ -1 }_D(p_2).\nonumber\\
\end{eqnarray}

\section{Matrix element of $\Lambda_b \rightarrow n l^+ l^- $ and $\Lambda_b \rightarrow n  \gamma $ decays}

In the standard model, the $\Lambda_b\rightarrow n l^+l^-$ transition is described by $b\rightarrow d l^+l^-$ at the quark level.
The effective Hamiltonian describing the electroweak penguin and weak box diagrams related to this transition is given by

\begin{eqnarray}
\mathcal{H}( b\rightarrow d l^+l^-) &=&  \frac{G_F\alpha}{2 \sqrt{2}\pi}V_{tb}V^*_{td}\bigg[ C^{eff}_9 \bar{d} \gamma_{\mu}(1-\gamma_5)b \bar{l}\gamma^{\mu}l  - i C^{eff}_{7}\bar{d}\frac{2 m_b\sigma_{\mu\nu} q^{\mu}}{q^2}(1+\gamma_5) b \bar{l}\gamma^{\mu}l  \nonumber \\
~~~~~~~~~~&+&C_{10} \bar{d}\gamma_{\mu}(1-\gamma_5) b \bar{l}\gamma^{\mu}\gamma_5l  \bigg],
\end{eqnarray}
where $G_F$ is the Fermi coupling constant, $\alpha$ is the fine structure constant at Z mass scale, $\epsilon^\nu$ is the polarization vector of photon, respectively.  $q$ is the total momentum of the lepton pair and $C_i^{eff}~(i=7,~9,~10)$ are the Wilson coefficients,
$C^{eff}_7=-0.313$, $C^{eff}_9=4.334$, $C_{10}=-4.669$\cite{JHEP10-118, PRD79-074007, EPJC40-565}.
The amplitude is obtained by sandwiching the effective Hamiltonian between the initial and final states.
The matrix element for $\Lambda_b \rightarrow n$ can be parameterized in terms of the FFs as the following:
\begin{eqnarray}\label{FFs:12}
 \langle n(P^\prime) | \bar{d}\gamma_{\mu}b | \Lambda_b(P)\rangle &=& \bar{u}_{n}(P^\prime)(g_1\gamma^\mu+ ig_2\sigma_{\mu\nu}p^{\nu}+g_3p_\mu)u_{\Lambda_b}(P),\nonumber\\
 \langle n(P^\prime) | \bar{d}\gamma_{\mu}\gamma_{5}b  | \Lambda_b(P)\rangle &= & \bar{u}_{n}(P^\prime)(t_1\gamma^\mu+it_2\sigma_{\mu\nu}p^{\nu}+t_3p^\mu)\gamma_5u_{\Lambda_b}(P),\nonumber\\
 \langle n (P^\prime) | \bar{d}i\sigma^{\mu\nu}q^{\nu}b | \Lambda_b(P)\rangle &= & \bar{u}_{n}(P^\prime)(s_1\gamma^\mu+is_2\sigma_{\mu\nu}q^{\nu}+s_3q^\mu)u_{\Lambda_b}(P),\nonumber\\
 \langle n (P^\prime)  |  \bar{d}i\sigma^{\mu\nu}\gamma_5q^{\nu}b | \Lambda_b(P)\rangle &= & \bar{u}_{n}(P^\prime)(d_1\gamma^\mu+id_2\sigma_{\mu\nu}q^{\nu}+d_3q^\mu)\gamma_5u_{\Lambda_b}(P),
\end{eqnarray}
 where $P^\prime$ and $P$ are the momenta of the neutron and $\Lambda_b$ respectively, $q=P-P^\prime$, $u_n$ and $u_{\Lambda_b}$ are the spinors of the initial and final baryons respectively, $g_i$, $t_i$, $s_i $, and $d_i $ ($i=1,2$ and 3) are the transition FFs which are Lorentz scalar functions of $q^2$.
When working in the limit $m_b\rightarrow \infty $, the number of independent FFs is reduced to $2$.
The $\Lambda_b \rightarrow n$ matrix element with an arbitrary matrix $\Gamma$ is given by
\begin{eqnarray}\label{FFs-HQET}
  \langle n (P^\prime)| \bar{d}\Gamma b\arrowvert \Lambda_b(v)\rangle =\bar{u}_{n}(P^\prime)(F_{1}(\omega)+F_2(\omega)\slashed{v})\Gamma u_{\Lambda_b}(v),
\end{eqnarray}
where $\Gamma= \gamma_\mu,~\gamma_\mu \gamma_5,~ q^\nu \sigma_{\nu\mu},q^\nu\sigma_{\nu\mu}\gamma_5$.
 $F_1$ and $F_2$  can be expressed as functions solely of $\omega=v\cdot P^\prime/m_{\Lambda}$, which is the energy of the neutron in the $\Lambda_b$ rest frame.
The baryons states can be normalized as follows,
\begin{eqnarray}
  \langle n(P^\prime)|n(P)\rangle &=& 2 E_n (2\pi)^3 \delta^3(P-P^\prime), \\
  \langle \Lambda_b(v^\prime,P^\prime)|\Lambda_b(v,P)\rangle &=& 2 v_0(2\pi)^3 \delta^3(P-P^\prime). \\
\end{eqnarray}

Comparing Eq. (\ref{FFs:12}) with Eq. (\ref{FFs-HQET}), we obtain the following relations:
\begin{eqnarray}
 & & g_1~=~t_1~=~s_2~=~d_2~=~\bigg(F_1+\sqrt{r}F_2\bigg),\nonumber\\
 & & g_2~=~t_2~=g_3~=~t_3~=~\frac{1}{m_{\Lambda_{b}}}F_2, \nonumber\\
 & & s_3~=~  F_2 (\sqrt{r}-1),~ d_3~=~ F_2(\sqrt{r}+1), \nonumber\\
 & & s_1 ~=~ d_1~=~ F_2 m_{\Lambda_b}  (1+r-2\sqrt{r}\omega),
\end{eqnarray}
where $r=m_n^2/m_{\Lambda_b}^2$.
On the other hand, the transition matrix for $\Lambda_b\rightarrow n$ can be expressed in terms of the BS wave functions of $\Lambda_b$ and $n$,
\begin{eqnarray}\label{FFs-BS}
  \langle n (P^\prime)|\bar{d}\Gamma b|\Lambda_b(P)\rangle =\int\frac{d^4p}{(2\pi)^4} \bar{\chi}_{P^\prime}^{n}(p^\prime)\Gamma \chi_P^{\Lambda_b}(p)S^{-1}_D(p_2).
\end{eqnarray}
where the $\chi_{P^\prime}^{n}$ and $\chi_{P^\prime}^{\Lambda_b}$ are the BS wave function of neutron and $\Lambda_b$ respectively.

Define
\begin{eqnarray}
  \int \frac{d^4p}{(2 \pi)^4} f_1(p^\prime) \phi(p) S^{-1}_D(p_2)&=&k_1(\omega), \nonumber\\
  \int \frac{d^4p}{(2 \pi)^4} f_2(p^\prime)p_{t\mu}^\prime \phi(p) S^{-1}_D(p_2)&=&k_2(\omega) v_{\mu} + k_3(\omega) v^\prime_{\mu},
\end{eqnarray}
where  $v^\prime = P^\prime/m_n$, then we find the following relations when $\omega \neq 1$:
\begin{eqnarray}
  k_3 &=& - \omega k_2, \nonumber\\
  k_2 &=& \frac{1}{1-\omega^2} \int \frac{d^4 p}{(2\pi)^4} f_2(p^\prime) p^\prime_t \cdot v \phi(p) S^{-1}_D, \nonumber \\
  F_1 &=& k_1- \omega k_2 , \nonumber\\
  F_2&=&k_2.
\end{eqnarray}

The differential decay rate of $\Lambda_b \rightarrow n l^+ l^-$ is obtained as:
\begin{eqnarray}
 \mathcal{M}(\Lambda_b\rightarrow n l^{+} l^{-})&=&\frac{G_F \lambda_t}{2\sqrt{2}\pi}  \big[\bar{l}\gamma_{\mu}l\{\bar{u}_{n}[\gamma_{\mu}(A_1+B_1+ (A_1-B_1)\gamma_5 ) \nonumber\\
 & +& i\sigma^{\mu\nu}p_{\nu}(A_2+B_2+ (A_2-B_2)\gamma_5 )]u_{\Lambda_b}\} \nonumber\\
&+&\bar{l}\gamma_{\mu}\gamma_5l\{\bar{u}_{n}[\gamma^{\mu}(D_1+E_1+ (D_1-E_1)\gamma_5 ) \nonumber \\
 &+&i\sigma^{\mu\nu}p_{\nu}(D_2+E_2+ (D_2-E_2)\gamma_5 )\nonumber\\
 &+&p^{\mu}(D_3+E_3+ (D_3-E_3)\gamma_5 )]u_{\Lambda_b}\}\big],
\end{eqnarray}
where $\lambda_t=|V_{tb}*V_{td}^*|$, the parameters $A_i$, $B_i$ and $D_j$, $E_j$ ($i=1,2$ and $j=1,2,3$) are defined as
\begin{eqnarray}
&&A_i=\frac{1}{2}\bigg\{C^{eff}_{9}(g_i-t_i)-\frac{2C^{eff}_7 m_b}{p^2}(d_i +s_i )\bigg\},\nonumber\\
& &B_i = \frac{1}{2}\bigg\{C^{eff}_{9}(g_i+t_i) - \frac{2C^{eff}_7m_b}{p^2}(d_i -s_i )\bigg\}, \nonumber\\
& &D_j = \frac{1}{2}C_{10}(g_j-t_j), ~E_j=\frac{1}{2}C_{10}(g_j+t_j).
\end{eqnarray}

In the physical region ($4m^2_l\leq q^2\leq (m_{\Lambda_b}-m_{n})^2$), the decay rate of $\Lambda_b\rightarrow n l^+l^-$ is obtained as

\begin{eqnarray}
\frac{d\Gamma(\Lambda_b\rightarrow n l^+l^-)}{dq^2}=\frac{G^2_F\alpha^2}{2^{13}\pi^5m_{\Lambda_b}} |V_{tb}V^*_{td}|^2v_l\sqrt{\lambda(1,r,s)} \mathcal{M}(s)  ,
\end{eqnarray}
where  $s= 1 +r  - 2 \sqrt{r} \omega $, $ \lambda(1,r,s)=1+r^2+s^2-2r-2s-2rs$, and  $v_l=\sqrt{1-\frac{4m^2_l}{s * m^2_{\Lambda_b}}}$, and the decay amplitude is given as \cite{EPJC45-151}

\begin{eqnarray}
   \mathcal{M}(s) &=& \mathcal{M}_0(s) +\mathcal{M}_2(s),
\end{eqnarray}
where
\begin{eqnarray}
  \mathcal{M}_0(s)&&=32m^2_l m^4_{\Lambda_b}s(1+r-s)(|D_3|^2+|E_3|^2) \nonumber\\
  &&64m^2_lm^3_{\Lambda_b}(1-r-s)Re(D^*_1E_3+D_3E^*_1)\nonumber\\
& &+64m^2_{\Lambda_b}\sqrt{r}(6m^2_l-M^2_{\Lambda_b}s)Re(D_1^*E_1)\nonumber\\
&& 64m^2_lm^3_{\Lambda}\sqrt{r}\big(2m_{\Lambda_b}s Re(D^*_3E_3) +(1-r+s)Re(D^*_1D_3+E^*_1E_3)\big)\nonumber\\
&&+32m^2_{\Lambda}(2m^2_l+m^2_{\Lambda}s)\bigg\{(1-r+s)m_{\Lambda_b}\sqrt{r}Re(A^*_1A_2+B^*_1B_2)\nonumber\\
& &-m_{\Lambda_b}(1-r-s)Re(A^*_1B_2+A^*_2B_1)  -2\sqrt{r}\big(Re(A^*_1B_1)+m^2_{\Lambda}s Re(A^*_2B_2)\big) \bigg \}\nonumber\\
& &+ 8 m^2_{\Lambda_b}\bigg[4m^2_l(1+r-s)+m^2_{\Lambda_b}((1+r)^2- s^2)\bigg](|A_1|^2+|B_1|^2)\nonumber\\
&&+8m^4_{\Lambda_b}\bigg\{4m^2_l[\lambda+(1+r-s)s]+m^2_{\Lambda_b}s[(1-r)^2-s^2]\bigg\}(|A_2|^2+|B_2|^2) \nonumber\\
& & - 8m^2_{\Lambda_b}\bigg\{4m^2_l(1+r-s)-m_{\Lambda_b}[(1-r)^2-s^2]\bigg\} (|D_1|^2+|E_1|^2) \nonumber\\
&&+ 8m^5_{\Lambda_b}sv^2\bigg\{-8m_{\Lambda_b}s\sqrt{r}Re(D^*_2E_2) +4(1-r+s)\sqrt{r}Re(D^*_1D_2+E^*_1E_2)\nonumber\\
&& -4(1-r-s) Re(D^*_1E_2+D^*_2E_1)+m_{\Lambda_b}[(1-r)^2-s^2] (|D_2|^2+|E_2|^2)\bigg\},
\end{eqnarray}
\begin{eqnarray}
  \mathcal{M}(s) &=& 8m^6_{\Lambda_b}s v_l^2\lambda(|A_2|^2+|B_2|^2+|C_2|^2+|D_2|^2) \nonumber\\ &- &8 m^4_{\Lambda_b}v_l^2\lambda(|A_1|^2+|B_1|^2+|C_1|^2+|D_1|^2).
\end{eqnarray}

Similarly, the Hamiltonian for exclusive rare radiative decay $\Lambda_b \rightarrow n \gamma $ with $\gamma$ as a real photon is given by
\begin{eqnarray}
\mathcal{H}( b\rightarrow d \gamma) =- \frac{i G_F e}{ 4 \sqrt{2}\pi^2}V_{tb}V^*_{td}C^{eff}_{7}\bigg[ m_b \bar{d} \sigma_{\mu\nu} q^{\mu} (1+\gamma_5) b  +m_d \bar{d} \sigma_{\mu\nu} q^{\mu} (1-\gamma_5) b  \bigg]\epsilon^\nu,
\end{eqnarray}
where $\epsilon^\nu$ is the polarization vector of the photon. Then, the decay width is given by
\begin{eqnarray}
  \Gamma(\Lambda_b \rightarrow n \gamma) = \frac{\alpha G_F^2 m_b^2 m_{\Lambda_b}^3}{2^6 \pi^4} |V_{tb} V^*_{td}|^2 |C_7^{eff}|^2[s_2^2(0) +d_2^2(0)] \bigg( 1-\frac{m_n^2}{m^2_{\Lambda_b}} \bigg)^3.
\end{eqnarray}

\section{Numerical analysis and discussion}

In this section we present a detailed numerical analysis of the rare decay $\Lambda_b \rightarrow n l^+ l^-$ and radiative decay  $\Lambda_b \rightarrow n \gamma$.
In our calculations, we take the masses of baryons as $m_{\Lambda_b}=5.62$ GeV, $m_n=0.94$ GeV \cite{PRD98-030001}, and the masses of quarks, $m_b=5.02$ GeV and $m_d=0.34$ GeV \cite{PRD95-054001, PRD87-076013, PRD91-016006}.
The variable $\omega$ varies from $1$ to $3.073,~3.069,~1.89$ for $e,~\mu,~\tau$, respectively.

\begin{figure}[!htb]
\begin{center}
\begin{minipage}[t]{0.45\linewidth}
 \includegraphics[width=6.5cm]{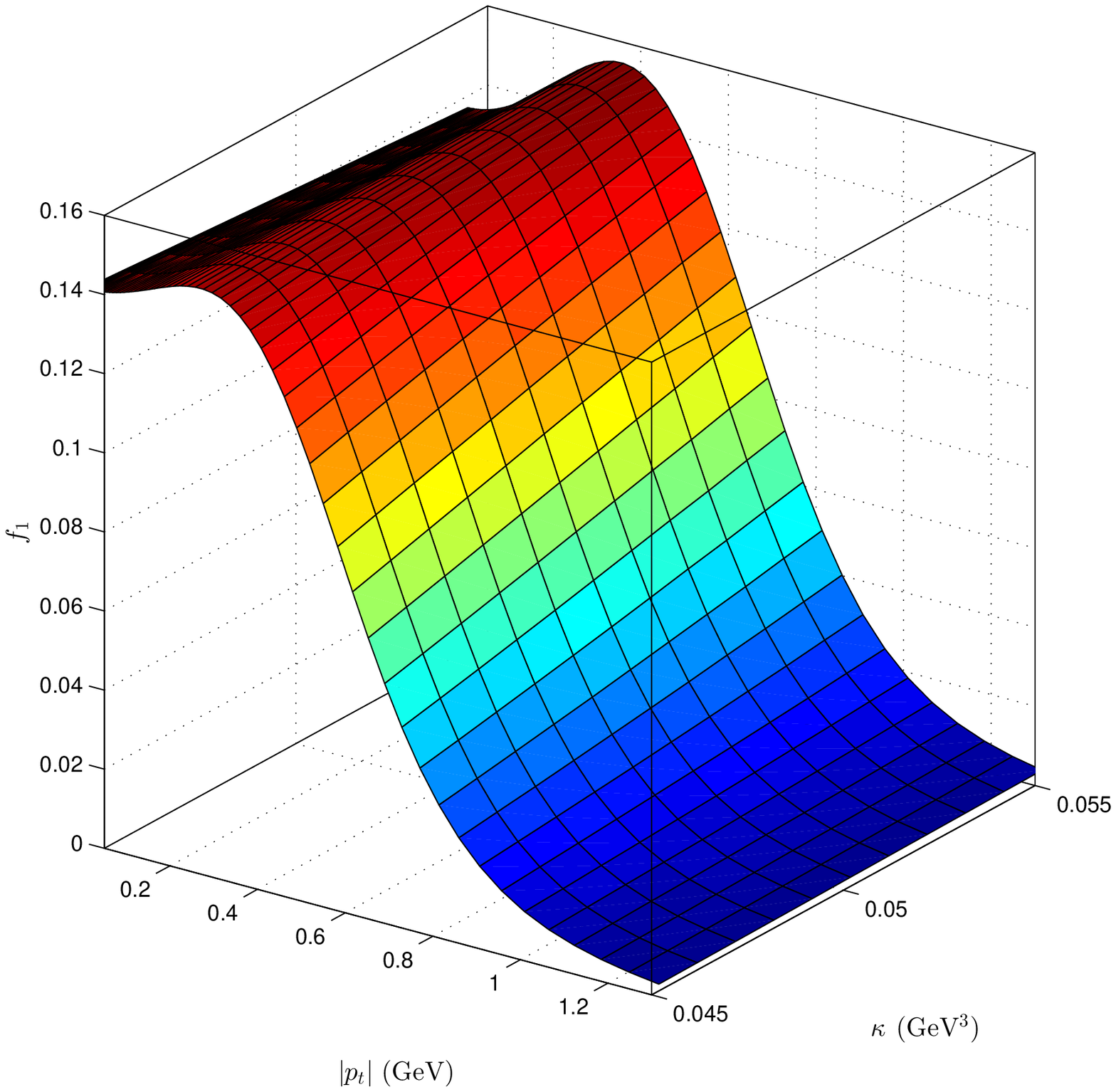}
\end{minipage}
\begin{minipage}[t]{0.45\linewidth}
 \includegraphics[width=6.5cm]{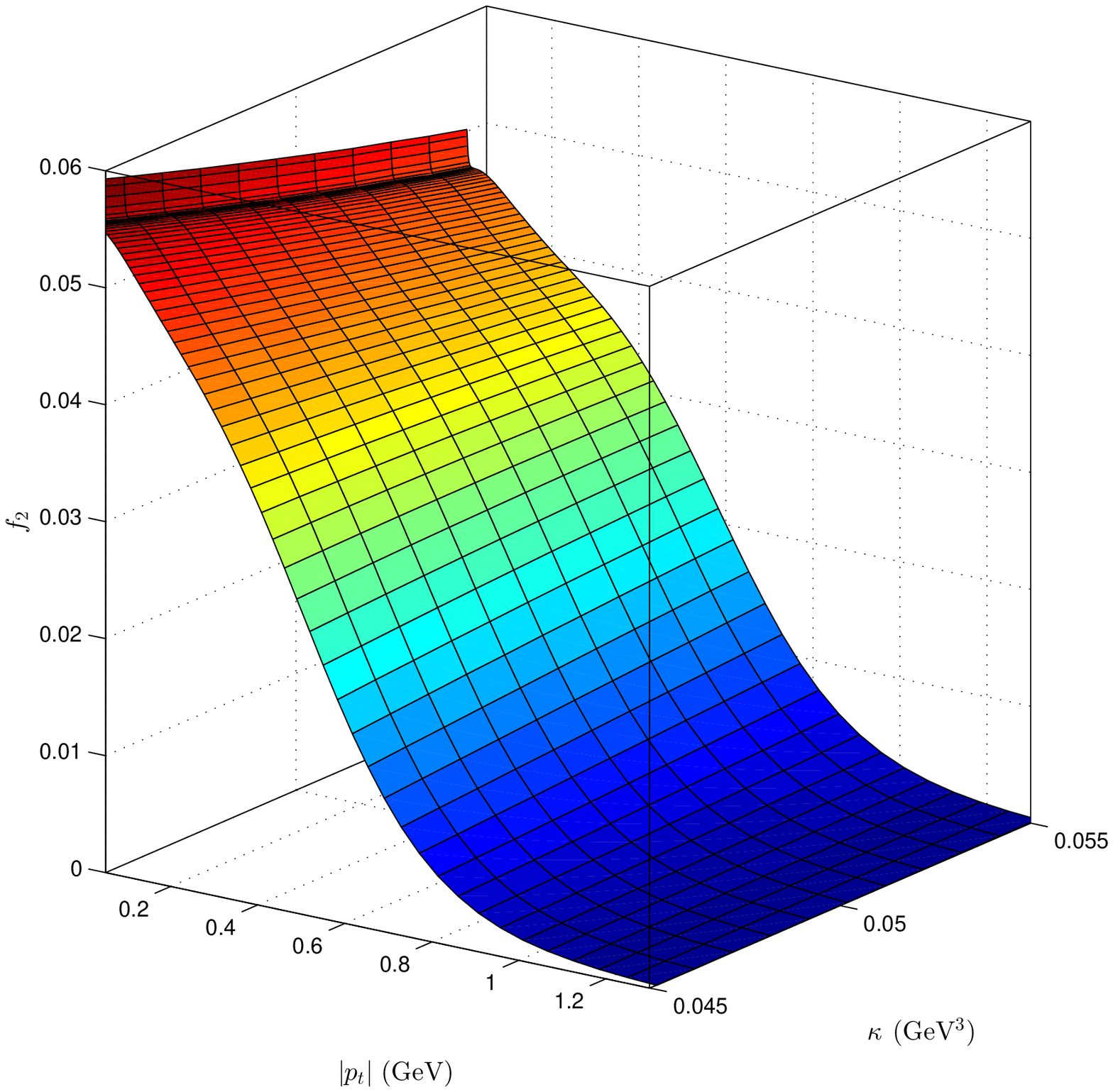}
\end{minipage}
\caption{(color online ) The BS wave functions for the neutron.}\label{Fig:n}
\end{center}
\end{figure}

\begin{figure}[!htb]
\begin{center}
 \includegraphics[width=6.5cm]{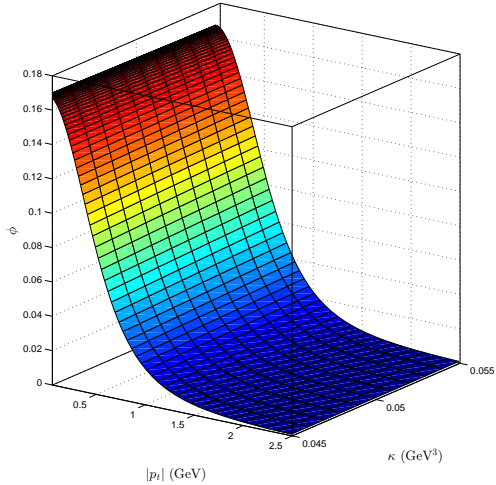}
\caption{(color online ) The BS wave function for $\Lambda_b$.}\label{Fig:b}
\end{center}
\end{figure}

Solving Eq. (\ref{BS:f1}), (\ref{BS:f2}) and (\ref{BS:hq}) for the neutron and $\Lambda_b$ with the parameters we have taken,  we get the numerical solutions of BS wave functions.
In Table. \ref{TB:alpha}, we give the values of $\alpha_s$ with different values of $\kappa $ for the neutron and $\Lambda_b$ and in Fig. \ref{Fig:n} and \ref{Fig:b}, we give the BS wave functions for the neutron and $\Lambda_b$.

\begin{table}[!htb]
\centering  
\begin{tabular}{c||c|c|c|c|c|c}  
\hline
  $\kappa$ (GeV$^3$)  & 0.045  & 0.047 &0.049 &0.051 &0.053  &0.055 \\ \hline \hline
 neutron &0.829 &0.811 & 0.793& 0.775& 0.758& 0.741 \\  \hline       
$\Lambda_b$  &0.775 & 0.777&0.778 &0.780& 0.782 &0.784 \\
 \hline
\end{tabular}
\caption{The values of $\alpha_{seff}$  for the neutron and $\Lambda_b$.}\label{TB:alpha}
\end{table}

\begin{figure}[!htb]
\begin{center}
\begin{minipage}[t]{0.45\linewidth}
 \includegraphics[width=6.50cm]{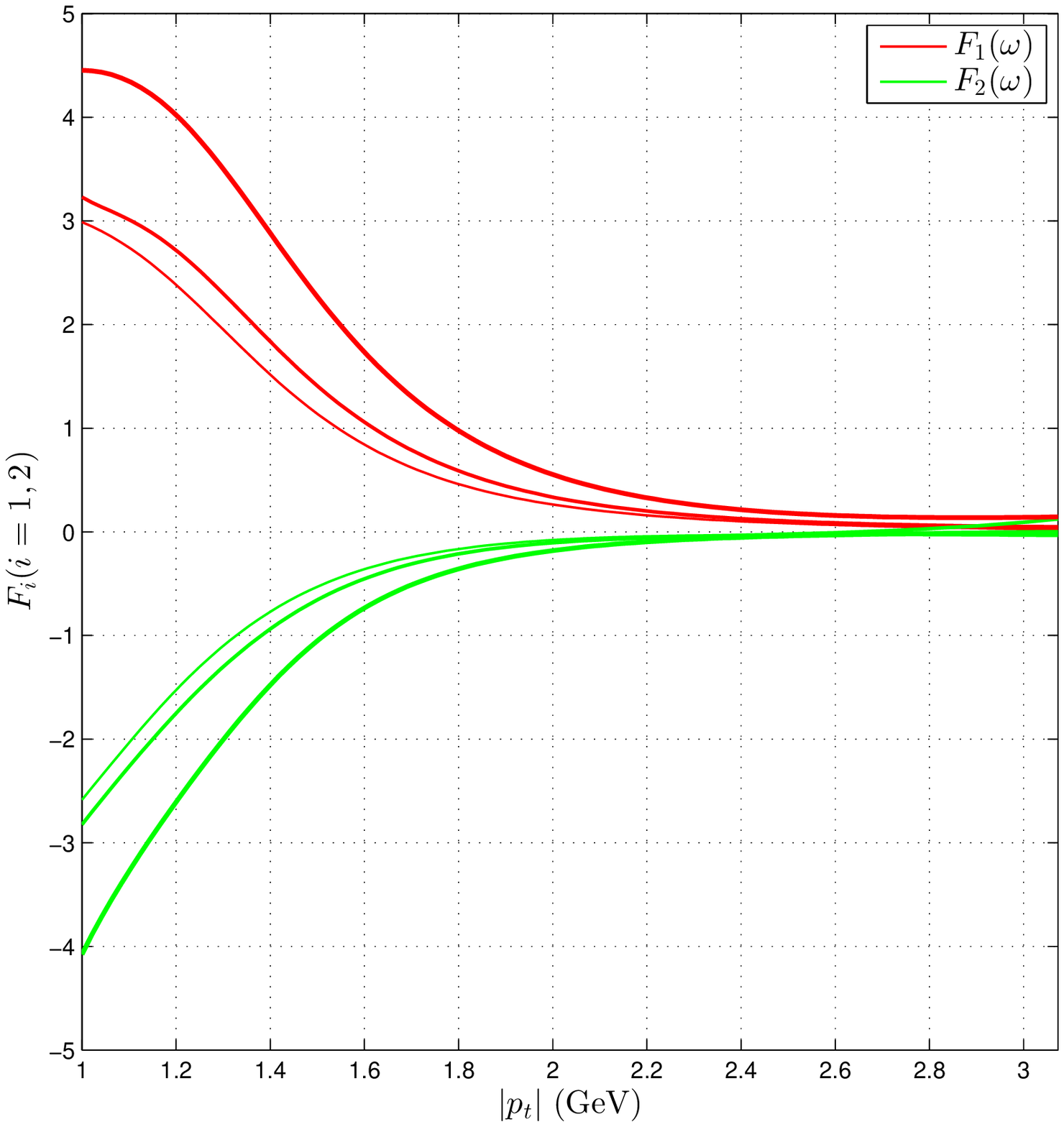}
\end{minipage}
\begin{minipage}[t]{0.45\linewidth}
 \includegraphics[width=6.50cm]{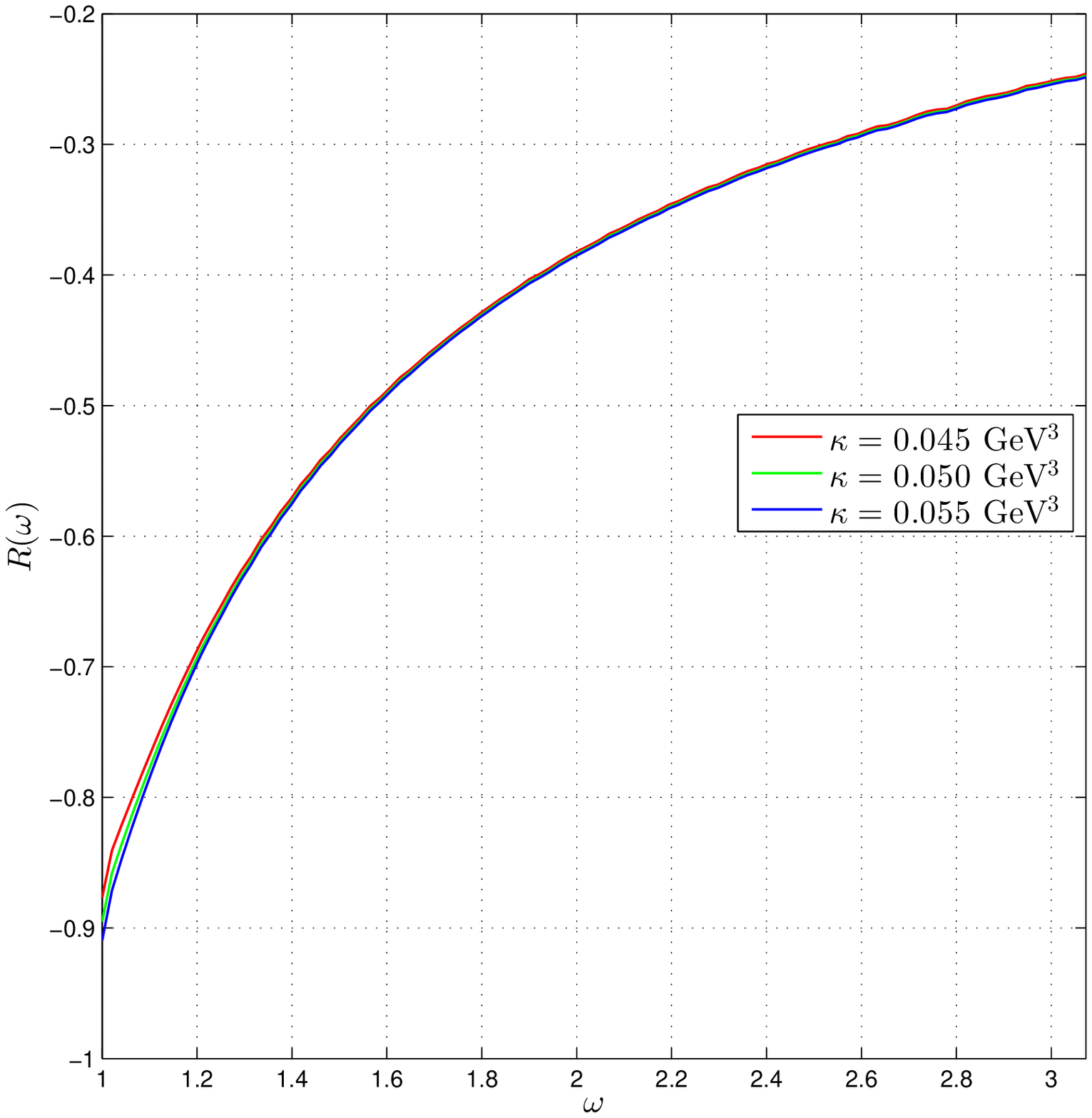}
\end{minipage}
\caption{(color online) The Values of FFs (the lines become thicker with the increases of $\kappa$) and $R(\omega)$ with different values of $\kappa$.}\label{FFs}
\end{center}
\end{figure}

It can be seen from Table \ref{TB:alpha} that the dependence of $\alpha_{seff}$ for the neutron on $\kappa$  is obviously stronger than that for $\Lambda_b$.
From the figures in Figs. \ref{Fig:n} and \ref{Fig:b}, we find that BS wave functions of neutron is very similarly on different $\kappa$, the values of $f_1(p_t)$ is about from $0$ to $0.14$ $f_2(p_t)$ varies about from $0$ to $0.06$ and $\phi(p_t)$ varies from $0$ to $0.17$.
In Fig. \ref{FFs}, we plot the FFs and $R(\omega)=F_2/F_1$ for different $\kappa$.
From this figure, we find that $F_1(\omega)$ increases with the increase of $\kappa$, but the value of $R(\omega)$ is not sensitive to the change of the value of $\kappa$.
The value of $R(\omega)$  varies from $-0.9$ to $-0.25$ when $\omega$ changes from $1$ to $3.1$.

In the heavy quark limit, assuming the same shape for $F_1$ and $F_2$, the ratio $R=-0.35\pm0.04$ (stat) $\pm0.04$ (syst) was previously  measured by the CLEO Collaboration using the experimental data for the semileptonic decay $\Lambda_c \rightarrow \Lambda e^+ \nu_e$ when $q^2$ changes from $m_\Lambda^2$ to $m^2_{\Lambda_c}$ \cite{PRL94-191801}.
In the same region, we find that $R(\omega)$ varies from $-0.32$ to $-0.25$ in our model.
In Ref. \cite{PRD59-114022} $R(\omega)$ varies from $-0.42$ to $-0.83$ when $q^2$ change from   $0$ to $(M_{\Lambda_b}-M_{\Lambda})^2$, and in our model $R(\omega)$  change from $-0.25$ to $-0.75$ in the same region.
However, in Ref. \cite{PRD53-4946} gives the behaviour $R(q^2)\propto -1/q^2$, which agrees with the pQCD scaling law \cite{PRD11-1309,PRD22-2157,PPNP59-694}.
 Therefore, using the CLEO Collaboration experimental data \cite{PRL94-191801}, we can estimate that the value of $R(\omega)$  should change from to $-0.91\pm0.03$ to $-0.3 \pm 0.03$ approximately, which agrees with our result as shown in Fig. \ref{FFs}.
From the data in Ref. \cite{JPGNPP37-115007}, we find that $R(\omega_{max})=-2.75 $  in LCSR and $R(\omega_{max})=-2.33$ by fit the data from LQCD \cite{PRL101-112002,PRD79-034504}.
From the data with the contribution of $\Lambda^*_b$  being considered Ref. \cite{PRD98-035033}, we find that $R(\omega_{max})=-3.47$ in LCSR.
These results are much larger than experimental data $R(\omega_{max})=-0.35$ \cite{PRL94-191801} and do not agree with our result.

\begin{table}[!htb]
\centering  
\begin{tabular}{c||c|c|c|c|c}  
\hline
 &  present work  & LCSR \cite{JPGNPP37-115007}  & LQCD \cite{JPGNPP37-115007} & LCSR \cite{PRD98-035033}  &  Ref. \cite{MPLA32-1750125} \\ \hline \hline
$Br( \Lambda_b\rightarrow n e^+ e^- ) \times 10^{8}$&$6.79_{-1.82}^{+8.66}$ &3.79$\pm$0.46 &3.19$\pm$0.32 &8$\pm$2 &3.81 \\         
$Br( \Lambda_b\rightarrow n \mu^+ \mu^- )\times 10^{8}$&$4.08^{+5.44}_{-1.19}$&3.76$\pm$0.42 &3.15$\pm$0.29 &7$\pm$2 &3.75 \\
$Br( \Lambda_b\rightarrow n \tau^+ \tau^- )\times 10^{8}$&$2.9^{+3.7}_{-0.78}$& 1.65$\pm$0.19&1.42$\pm$0.13 &2$\pm$0.4 &1.21  \\ \hline
$Br( \Lambda_b\rightarrow n \gamma)\times 10^{7}$ & $3.69^{+3.76}_{-1.95}$ & -&-&-&3.7\\
 \hline \hline
\end{tabular}
\caption{The values of the branching ratios of $\Lambda_b\rightarrow n l^+ l^-$ and $\Lambda_b\rightarrow n \gamma$ and compare with other model.}\label{TB3}
\end{table}

\begin{figure}[!htb]
  \centering
  \includegraphics[width=8.0cm]{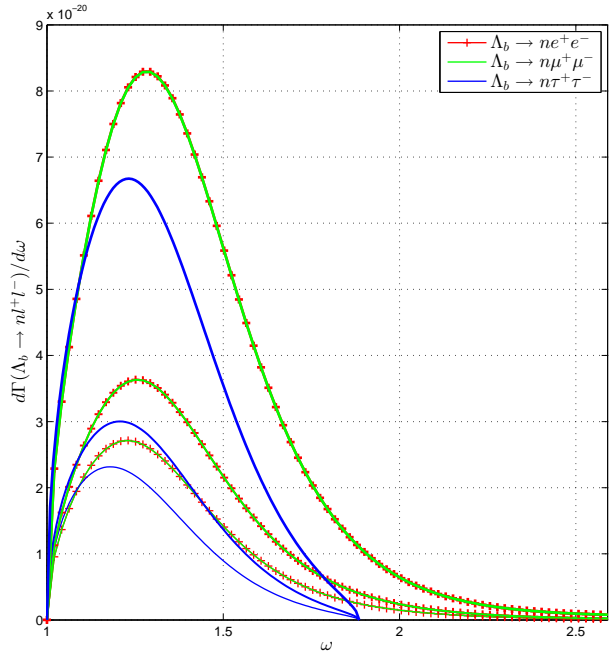}
  \caption{ (color online ) The decay widths of $\Lambda_b \rightarrow n l^+ l^-$  (the values of the decay width increase with the increase of $\kappa$ from $0.045$ to $0.055$) for the lines with the same color).}\label{DW}
\end{figure}

In Fig. \ref{DW}, we give the $\omega$-dependence of the decay widths of $\Lambda_b \rightarrow n l^- l^+(l= e, \mu, \tau)$ for different parameters.
For the central values of parameters, we find that the branching ratio are $BR(\Lambda_b \rightarrow n l^- l^+)\times10^8 = 6.79~(l=e),~4.08~(l=\mu),~2.90~(l= \tau)$ and $BR(\Lambda_b \rightarrow n \gamma)\times10^7 = 3.69$.
Our result for the branching ratios of $BR(\Lambda_b \rightarrow n l^- l^+)$ and $BR(\Lambda_b \rightarrow n \gamma)$ are listed in Table \ref{TB3} together those in other approaches.

Table \ref{TB3}, we can see that the rare semileptonic decay branching fractions are of order $10^{-8}$, and the rare radiative decay  branching fraction is of order $10^{-7}$.
In Ref. \cite{JPGNPP37-115007}, the authors use the parameters from LCSR \cite{PRD73-094019} and LQCD \cite{PRL101-112002, PRD79-034504} to fit the FFs of $\Lambda_b \rightarrow n$ and gave the branching ratio of $\Lambda_b \rightarrow n l^+ l^-$.
In Ref. \cite{PRD98-035033}, the authors also calculated the FFs of $\Lambda_b \rightarrow n$ in the framework of LCSR, but their results were different.
Our results for the branching ratios of $\Lambda_b \rightarrow n l^+ l^-~(l=e,~\tau)$ are very similar to those in Ref. \cite{PRD98-035033}, our result for $BR(\Lambda_b \rightarrow n \mu^+ \mu^-)$ agrees with that in Ref. \cite{JPGNPP37-115007}.
Our radiative decay result $BR(\Lambda_b \rightarrow n \gamma)$ agrees with Ref. \cite{MPLA32-1750125}.

\section{summary}

In our work, we calculated the FFs between baryons states induced by the rare $b \rightarrow d$ transition in the BS equation approach in a covariant quark-diquark model.
In our model, $\Lambda_b $ is regarded as a bound state of the $b$-quark and the scalar $ud$ diquark, thus only the $d^\uparrow (du)_{00}/ \sqrt{3}$ component of the neutron contributes to the FFs.
We established the BS equations for the $q(ud)_{00} ~(q=b,d)$ system and derived the FFs for $\Lambda_b \rightarrow n $ in the BS equation approach.
We solved the BS equation of $q(ud)_{00} ~(q=b,d)$ system and then we calculated the FFs and $R$ numerically.
Using these FFs, we obtained the branching ratios of $ \Lambda_b \rightarrow n l^+ l^-$ and $ \Lambda_b \rightarrow n \gamma$.
Comparing with other works we found that our FFs are very different with other model \cite{JPGNPP37-115007,MPLA32-1750125}, but the branching fractions of the semileptonic decay are of the order $10^{-8}$ and the radiative branching ratio is of the order $10^{-7}$.

In the near future, our results can be tested at LHCb.
Our model can be used to study the forward-backward asymmetries and CP violation in the rare decays of $b$ baryons to check our FFs.

\acknowledgments
This work was supported by National Natural Science Foundation of China under contract numbers 11775024,11575023,11847052,11981240361 and 11905117.


\end{document}